# Enabling Node Repair in Any Erasure Code for Distributed Storage

K. V. Rashmi, Nihar B. Shah, and P. Vijay Kumar, *Fellow, IEEE*

*Abstract*—Erasure codes are an efficient means of storing data across a network in comparison to data replication, as they tend to reduce the amount of data stored in the network and offer increased resilience in the presence of node failures. The codes perform poorly though, when repair of a failed node is called for, as they typically require the entire file to be downloaded to repair a failed node. A new class of erasure codes, termed as regenerating codes were recently introduced, that do much better in this respect.

However, given the variety of efficient erasure codes available in the literature, there is considerable interest in the construction of coding schemes that would enable traditional erasure codes to be used, while retaining the feature that only a fraction of the data need be downloaded for node repair. In this paper, we present a simple, yet powerful, framework that does precisely this. Under this framework, the nodes are partitioned into two *types* and encoded using two codes in a manner that reduces the problem of node-repair to that of erasure-decoding of the constituent codes. Depending upon the choice of the two codes, the framework can be used to avail one or more of the following advantages: simultaneous minimization of storage space and repair-bandwidth, low complexity of operation, fewer disk reads at helper nodes during repair, and error detection and correction.

## I. Introduction

In a distributed storage system, a data-file (the *message*) is dispersed across nodes in a network in such a manner that an end-user (also referred to as a *data-collector*) can retrieve the message by tapping into neighbouring nodes. It is desirable that a distributed storage system be reliable and use network resources such as storage and network bandwidth sparingly. The simplest means of increasing reliability of a storage system is through replication, i.e., by storing identical copies of the message in multiple storage nodes. However, for a given level of reliability, such systems are inefficient in storage space utilization as compared to other approaches.

A popular option that reduces the data stored in the network, and leads to increased resiliency in the face of node failures, is to employ erasure coding. Let the message to be stored in the network be represented by a collection of $B$ message symbols, with each message symbol drawn from a finite field $\mathbb{F}_q$ of size $q$. With (MDS) erasure codes, data is stored across $n$ nodes in the network in such a way that the entire message can be recovered by a data-collector by connecting to any arbitrary $k$ nodes. This process of data recovery is referred to as *data-reconstruction*. Several distributed storage systems such as RAID-6, OceanStore and Total Recall employ such an erasure-coding option.

The authors are with the Dept. of ECE, Indian Institute Of Science, Bangalore, India. Email: {rashmikv, nihar, vijay}@ece.iisc.ernet.in. P. Vijay Kumar is also an adjunct faculty member of the Electrical Engineering Systems Department at the University of Southern California, Los Angeles, CA 90089-2565.

Another important aspect of distributed storage system design is the handling of node failures. Upon failure of an individual node, a self-sustaining data storage network must necessarily possess the ability to *repair* (or, *regenerate*) the failed node. A typical means of accomplishing this task under erasure coding, is to first permit the replacement node to download the entire data stored in any $k$ nodes and then proceed to extract the data that was stored in the failed node. Such a procedure is clearly wasteful of network resources. This raises a natural question as to whether there is a better alternative. Such an alternative is provided by the concept of a *regenerating code* introduced in the pioneering paper by Dimakis et al. [1].

In the regeneration framework introduced in [1], codes whose symbol alphabet is a vector over $\mathbb{F}_q$, i.e., an element of $\mathbb{F}_q^\alpha$ for some integer parameter $\alpha > 1$ are employed. Given the vector nature of the code-symbol alphabet, we may equivalently, regard each node as storing a collection of $\alpha$ symbols, each symbol drawn from $\mathbb{F}_q$. Apart from this new parameter $\alpha$, two other parameters $(d, \beta)$ are associated with this framework. A replacement node is permitted to connect to an arbitrary subset of $d$ ($\geq k$) nodes out of the remaining $(n-1)$ nodes and download $\beta \leq \alpha$ symbols from each. These $d$ nodes helping in the repair of a failed node are termed as *helper nodes*. The total amount $d\beta$ of data downloaded for repair purposes is termed the *repair-bandwidth*. Typically, with a regenerating code, the average repair-bandwidth $d\beta$ is small compared to the size $B$ of the message.

A major result in the field of regenerating codes is the proof in [1] that uses the cut-set bound of network coding to establish that the parameters of a regenerating code must satisfy:

$$B \leq \sum_{i=0}^{k-1} \min\{\alpha, (d-i)\beta\} \ . \qquad (1)$$

It can be deduced (see [1]) that achieving equality in (1), with parameters $B$, $k$ and $d$ fixed, leads to a tradeoff between the storage space $\alpha$ and the repair-bandwidth $d\beta$. This tradeoff is termed as the storage vs repair-bandwidth tradeoff.

The property of reconstruction of the $B$ source symbols from $k$ nodes requires the storage per node $\alpha$ to be at least $\frac{B}{k}$. The case when $\alpha = \frac{B}{k}$ is termed the Minimum Storage Regeneration (MSR) point. On the other hand, to repair a failed node storing $\alpha$ symbols, one must necessarily download at least $\alpha$ symbols from the network, and the case when the repair bandwidth $d\beta$ is equal to $\alpha$ is termed the Minimum Bandwidth Regeneration (MBR) point. However, both these parameters cannot be minimized simultaneously: the minimum storage case requires a repair bandwidth of $d\beta \geq \alpha + (k-1)\beta$ while the minimum bandwidth point requires a storage per node of $\alpha \geq \frac{B}{k} + \frac{(k-1)}{2}\beta$.

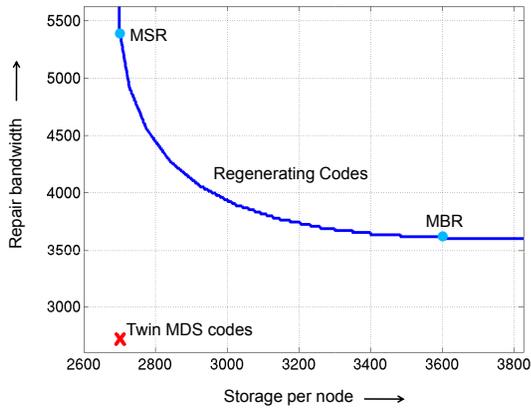

Fig. 1: The storage vs repair-bandwidth tradeoff in the regenerating codes setup, when $B = 27000$, $k = 10$, $d = 18$, and $n > 18$. The Twin-code framework, by relaxing certain constraints in the regenerating codes setup, can operate at the point (storage = 2700, repair-bandwidth = 2700).

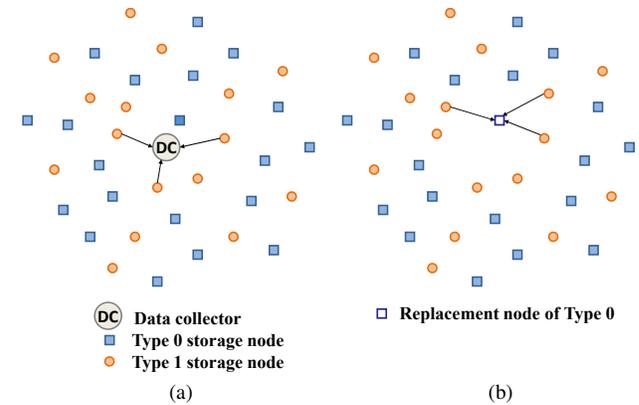

Fig. 2: The setting: (a) a data-collector connecting to a subset of storage nodes of Type 1, and (b) a replacement node of Type 0 connecting to a subset of storage nodes of the opposite type, Type 1.

Fig. 1 plots the tradeoff for the case when $B = 27000$ symbols, $k = 10$, and $d = 18$. For this example parameter set, the minimum possible storage space required at each node can be computed to be 2700 symbols, for which the repair-bandwidth required is at least 5400 symbols. On the other hand, at the MBR point, the minimum possible repair-bandwidth is 3600 symbols, while each node is required to store at least 3600 symbols.

For an overview of the explicit codes and schemes under the regenerating codes framework, non-existence results, and alternative models, the reader is referred to [2], [3] and the references therein.

In the present paper, we propose a novel, simple, yet powerful, 'Twin-code framework' for a distributed storage network that facilitates both data-reconstruction and efficient node-repair. The framework attempts to answer the following two questions:

*Q1.* As seen above, it is not possible to simultaneously minimize both repair-bandwidth and storage per node in the regenerating codes setup. On the other hand, in many applications, both storage and bandwidth may be expensive, while the property that data-reconstruction and node-repair be possible from *any* subset of nodes of size $k$ and $d$ respectively may not be required. A natural question that arises is whether it is possible to further reduce storage and repair-bandwidth by relaxing these constraints (while still retaining the property of having several subsets of nodes to choose from during reconstruction and repair).

*Q2.* The rich history of research on erasure codes makes available a wide variety of codes with various useful properties. Thus, another question that arises is whether these codes can be employed in a distributed storage network while still enjoying the benefits of efficient node repair.

Recently, there has been some work ([4], [5]) in the literature that relate to the questions posed above. In [4], the authors relax certain requirements on node-repair to generalize the repair-by-transfer (uncoded-repair) MBR codes constructed in [6] to a wider range of parameters. In [5], the authors provide mechanisms to lower the repair-bandwidth in certain existing Array codes.

Under the Twin-code framework introduced in the present paper, nodes in the network are partitioned into two types, labeled as Type 0 and Type 1 nodes. To arrive at the data stored in nodes of Type 0, the message is encoded using a linear code $C_0$. To arrive at the data stored in Type 1 nodes, the message symbols are first permuted by means of simple transposition and then encoded by a second linear code $C_1$. The two codes do not necessarily have to be distinct. Repair of a failed node of one type is accomplished by downloading data from a subset of nodes of the *other* type. Further, a data-collector can recover the entire message by connecting to a subset of the nodes of the *same* type (see Fig. 2).[1]

Under this framework, the operations of encoding, data-reconstruction and node-repair employ the encoding and decoding algorithms of the constituent codes. And it is this feature we feel, that makes the approach powerful. The characteristics of the resulting network reflect features such as erasure performance and decoding complexity of the two constituent codes. For example:

(a) when the two constituent codes are chosen to be Maximum-Distance-Separable (MDS), for example Reed-Solomon codes, both storage per node as well as repair-bandwidth are minimized simultaneously (see Fig. 1).

(b) when the two constituent codes are chosen to be codes such as LDPC or Fountain codes that possess very low erasure decoding complexity, the result is a distributed storage network in which efficient data-reconstruction and node-repair can be carried out in low-complexity fashion.

(c) when codes possessing sparse generator matrices are employed, the repair can be carried out with very few disk reads at the helper nodes.

(d) when the two constituent codes are chosen to be MDS Array codes, such as EVENODD codes, they minimize both resources, as well as lower the complexity to a certain extent. The latter is by virtue of the property of Array codes that they operate in the binary field and employ XOR operations alone.

The Twin-code framework, as well as a specialization to the case when the constituent codes are MDS, is provided in the next section, Section II. The principal advantages of this framework are presented in Section III.

---

[1]The subsets of nodes to which a data-collector or replacement node can connect to are governed by the decoding algorithms of the constituent codes $C_0$ and $C_1$. This is made clear in Section II.

## II. THE TWIN-CODE FRAMEWORK

Under the Twin-code framework, the $n$ storage nodes in the network are partitioned into two categories, consisting of $n_0$ and $n_1$ nodes, which we will refer to as *Type 0* and *Type 1* nodes respectively. The message to be stored across the network comprises of $B$ symbols drawn from the finite field $\mathbb{F}_q$. For $i=0,1$, let $\mathcal{C}_i$ be an arbitrary $[n_i, k]$ linear code over $\mathbb{F}_q$ having $(k \times n_i)$ generator matrix $G_i$. Let $\underline{g}_{(i,\ell)}$ $(1 \leq \ell \leq n_i)$ denote the $\ell^{th}$ column of $G_i$.

### A. Encoding

The message is first divided into fragments of $k^2$ symbols.[2] The rest of the section describes the code for a single fragment, and constructions for the case when the size of the message is larger can be obtained by concatenating multiple such fragments. Thus, we have
$$B = k^2. \qquad (2)$$
Let the $k^2$ symbols be arranged in the form of a $(k \times k)$ matrix $M_0$, which we will call the *message matrix*. Let
$$M_1 \triangleq M_0^t, \qquad (3)$$
where the superscript '$t$' denotes the transpose. For $i=0,1$, the data stored in the nodes of Type $i$ is obtained using code $\mathcal{C}_i$ and message matrix $M_i$ as follows. Each node of Type $i$ stores the $k$ symbols corresponding to a column of the $(k \times n_i)$ matrix $M_i G_i$. More specifically, node $\ell$ $(1 \leq \ell \leq n_i)$ of Type $i$ stores the symbols in its $\ell^{th}$ column (see Fig. 3), given by
$$M_i \underline{g}_{(i,\ell)}. \qquad (4)$$
Thus, under our encoding algorithm, every node stores $k$ symbols.[3] Also, each node $\ell$ of Type $i$ is associated with a distinct column $\underline{g}_{(i,\ell)}$ of $G_i$, and we will refer to $\underline{g}_{(i,\ell)}$ as the *encoding vector* of this node.

This completes the description of how the data is encoded and mapped onto the network. In the sections to follow, we first discuss data-reconstruction and node-repair for the case when both $\mathcal{C}_0$ and $\mathcal{C}_1$ are MDS codes, before moving on to the general case. MDS codes serve as a simple and concrete example, and also turn in a strong performance by minimizing both storage per node and repair-bandwidth. We call the codes resulting from the Twin-code framework when the constituent codes are MDS as *Twin-MDS (distributed storage) codes*.

### B. Twin-MDS Codes

In this subsection, we specialize to the case when both $\mathcal{C}_0$ and $\mathcal{C}_1$ are MDS codes over $\mathbb{F}_q$. In this case:
- a data-collector can recover the entire message by connecting to *any* $k$ nodes of the same type and downloading the $k^2$ symbols stored in them. Note that a connectivity of $2k-1$ or higher guarantees the availability of such a subset.

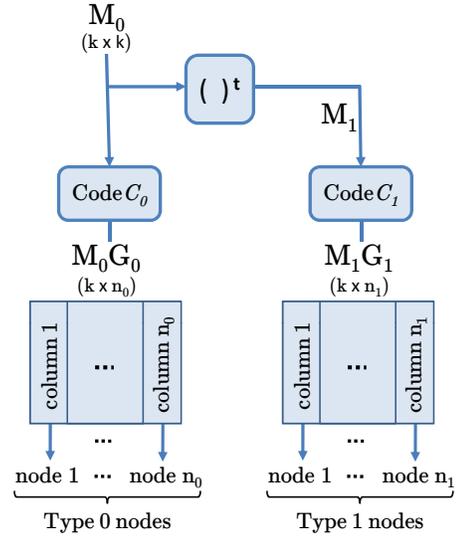

Fig. 3: The encoding procedure under the Twin-code framework.

- a replacement node of a certain type can recover the $k$ symbols that were stored in the failed node by downloading a single symbol over $\mathbb{F}_q$ from *any* subset of $k$ nodes belonging to the other type.[4]

To see how data-reconstruction is accomplished in this case, let us assume without loss of generality that the data-collector connects to the first $k$ nodes of Type 1. The data-collector gains access to the $k^2$ symbols given by the matrix:
$$M_1 \begin{bmatrix} \underline{g}_{(1,1)} & \cdots & \underline{g}_{(1,k)} \end{bmatrix}. \qquad (5)$$
The MDS property of code $\mathcal{C}_1$ guarantees linear independence of the corresponding $k$ columns of $G_1$. Thus, the data-collector can recover the message matrix $M_1$ by inverting the matrix formed by these columns.

Equivalently, the action of the data-collector can also be viewed as erasure decoding of code $\mathcal{C}_1$ in the presence of $(n_1 - k)$ erasures. To see this, consider the $k$ symbols in the first row of the matrix in (5). Clearly, these are $k$ components of the codeword of code $\mathcal{C}_1$ corresponding to the $k$ message symbols in the first row of $M_1$. Thus, each row of the message matrix $M_1$ can be decoded independently, allowing the data-collector to perform scalar decoding of the vector code.

To see how node-repair is accomplished in this case, let us assume that node $f$ of Type 0 fails. Thus the replacement node needs to recover the $k$ symbols
$$M_0 \underline{g}_{(0,f)}. \qquad (6)$$
We further assume without loss of generality that it connects to the first $k$ nodes of Type 1. Under the Twin-code framework, the replacement node makes known its identity to each of the helper nodes. In turn, the $\ell^{th}$ helper node $(1 \leq \ell \leq k)$ passes the inner product of the $k$ symbols $M_1 \underline{g}_{(1,\ell)}$ stored in it with the encoding vector $\underline{g}_{(0,f)}$ of the failed node: $\underline{g}_{(0,f)}^t M_1 \underline{g}_{(1,\ell)}$. Thus, the replacement node gains access to the $k$ symbols
$$\underline{g}_{(0,f)}^t M_1 \begin{bmatrix} \underline{g}_{(1,1)} & \cdots & \underline{g}_{(1,k)} \end{bmatrix}. \qquad (7)$$

---

[2] In the terminology of distributed storage, each fragment is called a *stripe*.
[3] It can be shown that the encoded symbols correspond to the first and the third quadrants of a Product-code [7] built out of the constituent codes. For lack of space, we do not elaborate here.

[4] Regenerating codes permit node-repair using *any* $d$ ($\geq k$) nodes in the system. However for a setting where storage per node is to be minimized (MSR), the value of $d$ must be much larger than $k$ to make the repair-bandwidth close to the fraction of data stored per node.

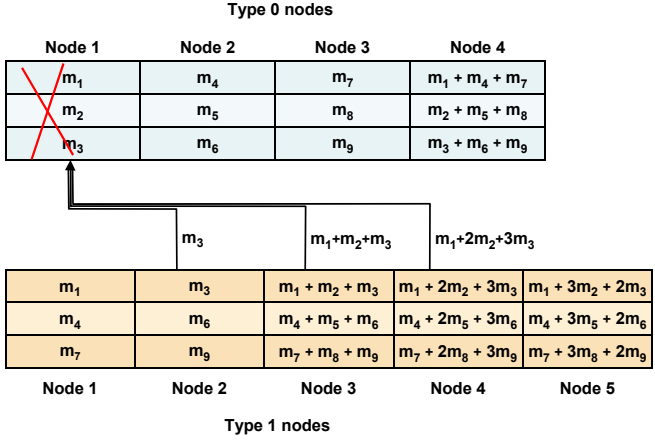

Fig. 4: An example of the Twin-code framework with $\mathcal{C}_0$ and $\mathcal{C}_1$ as MDS codes with parameters $n_0 = 4$, $n_1 = 5$ and $k = 3$.

Now, setting
$$\underline{\mu}^t \triangleq \underline{g}^t_{(0,f)} M_1 , \qquad (8)$$
we see that the replacement node has access to
$$\underline{\mu}^t \left[ \underline{g}_{(1,1)} \cdots \underline{g}_{(1,k)} \right] . \qquad (9)$$
It is clear that $\underline{\mu}^t$ can be recovered by erasure decoding of the MDS code $\overline{\mathcal{C}}_1$ under $(n_1 - k)$ erasures. Recall that by construction, we have $M_0 = M_1^t$, and hence
$$\underline{\mu} = (\underline{g}^t_{(0,f)} M_1)^t = M_0 \, \underline{g}_{(0,f)} . \qquad (10)$$
Thus, the vector $\underline{\mu}$ comprises of precisely the $k$ symbols required by the replacement node. It follows that repair of a node of Type 0 is reduced to erasure decoding of code $\mathcal{C}_1$.

Fig. 4 illustrates a numerical example, where the network has $n = 9$ nodes, with $n_0 = 4$, $n_1 = 5$, $k = 3$, giving $B = k^2 = 9$. The code operates over $\mathbb{F}_7$. The message matrices $M_0$ and $M_1(= M_0^t)$ populated by the message symbols $\{m_i\}_{i=1}^9$, and the generator matrices $G_0$ and $G_1$ are:
$$M_0 = \begin{bmatrix} m_1 & m_4 & m_7 \\ m_2 & m_5 & m_8 \\ m_3 & m_6 & m_9 \end{bmatrix}, G_0 = \begin{bmatrix} 1 & 0 & 0 & 1 \\ 0 & 1 & 0 & 1 \\ 0 & 0 & 1 & 1 \end{bmatrix}, G_1 = \begin{bmatrix} 1 & 0 & 1 & 1 & 1 \\ 0 & 0 & 1 & 2 & 3 \\ 0 & 1 & 1 & 3 & 2 \end{bmatrix}.$$

It can be verified that this code permits a data-collector to reconstruct the entire message from any three nodes of the same type. The figure also depicts repair of node 1 of Type 0, which has an encoding vector $[1 \, 0 \, 0]^t$, with the help of nodes 2, 3 and 4 of Type 1.

### C. The Twin-Code Framework in a General Setting

In this subsection, we describe the Twin-code framework applied to arbitrary linear block codes. By a linear code we mean a code where every code symbol can be expressed as a linear combination of the message symbols. We consider each symbol to belong to the finite field $\mathbb{F}_q$. Data is encoded onto the network as described in Section II-A, and it remains to explain how data-reconstruction and node-repair are accomplished. We first define the term *feasible set of nodes* which will aid in the description.

*Definition 1 (Feasible set of nodes):* A feasible set of nodes is a set of nodes of the *same* type, say Type $i$, such that the columns of the generator matrix $G_i$ associated to these nodes permit erasure decoding of the code $\mathcal{C}_i$.

For example, when code $\mathcal{C}_0$ is an $[n_0, k]$ MDS code, any $k$ columns of its generator matrix $G_0$ are linearly independent, enabling erasure decoding from any $k$ columns of $G_0$. It follows that when $\mathcal{C}_0$ is MDS, any set of $k$ nodes of Type 0 is a *feasible set*.

Under the Twin-code framework, a data-collector can recover the entire message by connecting to *any* feasible subset of nodes of the *same* type. For example, when the constituent codes are MDS as in Section II-B, the message can be recovered from any $k$ nodes of the same type.

To see how data-reconstruction is accomplished, let us suppose that the data-collector connects to a feasible subset of Type 1 nodes. Restricting our attention to nodes of Type 1 alone, the system is identical to a distributed storage system employing only code $\mathcal{C}_1$. Thus, the data-collector can recover the entire message by applying the decoding procedure of code $\mathcal{C}_1$, $k$ times (once for each row of the message matrix $M_1$).

We now turn our attention to node-repair. Under the Twin-code framework, a replacement node of a certain type can recover the symbols stored in the failed node by downloading a single symbol from *any* feasible subset of the *other* type. Our repair algorithm reduces the problem of node-repair to one of data-reconstruction, however with an amount of download which is close to a small fraction of the entire data.

To see how node-repair is accomplished, let us assume failure of node $f$ of Type 0. The replacement node desires to recover the $k$ symbols $M_0 \, \underline{g}_{(0,f)}$. The replacement node connects to a feasible set $\mathcal{F}$ of Type 1 nodes. As in the case of Twin-MDS codes, each helper node $\ell \in \mathcal{F}$ passes the symbol $\underline{g}^t_{(0,f)} M_1 \, \underline{g}_{(1,\ell)}$ to the replacement node. Setting $\underline{\mu}^t \triangleq \underline{g}^t_{(0,f)} M_1$, we see that the replacement node gains access to the symbols $\{ \underline{\mu}^t \underline{g}_{(1,\ell)} \mid \ell \in \mathcal{F}\}$. Since the set of helper nodes form a feasible set of Type 1 nodes, it is clear that the replacement node can recover $\underline{\mu}^t$ through erasure decoding of code $\mathcal{C}_1$. Under our framework, $M_0 = M_1^t$ and hence, $\underline{\mu} = (\underline{g}^t_{(0,f)} M_1)^t = M_0 \, \underline{g}_{(0,f)}$. The vector $\underline{\mu}$ thus comprises of precisely the $k$ symbols required by the replacement node. In this manner, our repair algorithm requires a download equal to a fraction $\frac{1}{k}$ of the data download required during the data-reconstruction process.

*Remark 1 (Application to Vector Codes):* One could also employ vector codes such as EVENODD and other Array codes in the the Twin-code framework. Under a constituent code where each code symbol is a vector of length $\nu$, each storage node is associated to $\nu$ encoding vectors, specifically, the $\nu$ columns associated to a (vector) code symbol in the generator matrix of the constituent code. We do not delve into the details of such constructions here due to lack of space.

*Worst-case analysis:* In the Twin-MDS code, the system can clearly handle failure of any $n - (2k-1)$ nodes with respect to data-reconstruction. Optimal repair of a node, say of Type 0, can be performed even if any $(n_1 - k)$ nodes of Type 1 fail. In the situation when only $\hat{k} < k$ nodes of Type 1 are available, repair can be carried out with a greater download (assuming at least $k$ nodes of Type 0 are available) by first repairing $(k - \hat{k})$ (imaginary) nodes of Type 1, and then using this augmented collection of Type 1 nodes for repair. A similar worst-case analysis can be carried out when the constituent codes are not MDS, by replacing the parameter $k$ by $(n - d_{\min} + 1)$, where $d_{\min}$ is the minimum distance of the constituent code.

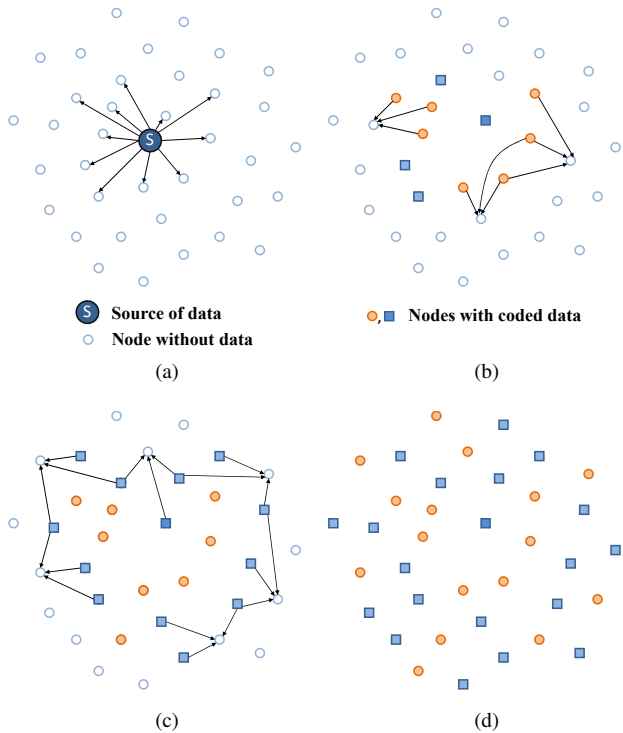

Fig. 5: Data deployment using the Twin-code framework: (a) The source transfers coded data to a subset of nodes. (b),(c) These nodes now help in the 'repair' of nodes in their neighbourhood that have not received data. The system eventually reaches a state as in (d).

## III. ADVANTAGES OF THE TWIN-CODE FRAMEWORK

### A. Employing Existing Codes

One of the main attractions of the Twin-code framework is that it permits the utilization of *any* linear erasure code. Moreover, the process of node-repair is reduced to one of erasure decoding of the constituent codes. Thus, this framework also allows a system designer to use the encoders and decoders of the constituent codes for all operations including node-repair.

### B. Reduced Repair Overhead

Under the Twin-code framework, the node-repair algorithm is such that the symbols passed by a helper node are independent of the identity of the other nodes helping in the repair process. Hence a helper node needs to be cognizant of only the encoding vector of the replacement node, allowing repair to be performed in an entirely distributed manner. This makes the implementation of such a system easier, and further reduces the communication overhead.

### C. Data Deployment

Codes with efficient node-repair properties, such as Regenerating codes and the Twin-code framework, can be employed for efficient deployment of data across a storage network. In such a scenario, the source first transmits the encoded data to a subset of nodes, following which, the source is no longer required to be available. Now, the nodes that have not received the data are treated as replacement nodes, and the repair algorithm is used to transmit data to this node (see Fig. 5 for an illustration). The distributed nature of such a deployment process will potentially make traffic more uniform across the network, thus aiding in load-balancing.

Twin-MDS codes possess the additional advantage of having the minimum possible repair-bandwidth of $\frac{B}{k}$. Thus, unlike any regenerating code, it reduces the total amount of data transferred during the deployment process to the minimum possible, and equal to the amount of data transferred when source directly transmits all encoded data.

### D. Error correction and detection

Although the preceding sections describe data-reconstruction and node-repair in the absence of errors, it is easy to see that for error-prone networks, one can use appropriate error correcting codes as the constituent codes in the Twin-code framework. The decoding algorithms of these codes can be employed for data-reconstruction and node-repair in the error-prone network.

### E. Extensions to the Twin-Code Framework

The Twin-code framework, when applied to two different constituent codes, allows the system designer to cater to two possibly different classes of end-users. For instance, the choice of $\mathcal{C}_0$ as a Reed-Solomon code and $\mathcal{C}_1$ as an LDPC code permits the bandwidth-constrained data-collector to connect to Type $0$ nodes and download minimum amount of data, and the computation-constrained data-collector to connect to Type $1$ nodes and perform data reconstruction with a low-complexity algorithm. In general, the Twin-code framework can be extended to include more than two types of nodes (representing the message as an array having a dimension equal to the number of types). Furthermore, one can choose constituent codes of different dimensions (instead of an identical dimension $k$), in which case, the message matrix will be rectangular. Finally, the alphabet need not be restricted to a finite field; any alphabet with two operators satisfying commutativity and distributivity can be employed, for example, a commutative ring.